\documentclass[a4paper,twocolumn,11pt,accepted=2022-09-22]{quantumarticle}
\pdfoutput=1

\usepackage{amsmath, amsfonts, amssymb}
\usepackage{hyperref}

\usepackage{graphicx}% Include figure files
\usepackage{dcolumn}% Align table columns on decimal point
\usepackage{bm}% bold math
\usepackage{braket}
\usepackage{xspace}
\usepackage[utf8]{inputenc} \usepackage[T1]{fontenc}

\usepackage{algorithm}
\usepackage{algpseudocode}
\renewcommand{\vec}[1]{ {\ensuremath{\bm{#1}} } }

\newcommand{\abs}[1]{\left|#1\right|}
\newcommand{\iden}{\mathbb{I}}
\newcommand{\oprod}{\bigotimes}
\newcommand{\order}[1]{\ensuremath{\mathcal{O}\left({#1}\right)}}
\newcommand{\etal}{\textit{et al}.\xspace}

\begin{document}

\title{Variational solutions to fermion-to-qubit mappings in two spatial dimensions}

\author{Jannes Nys}
\affiliation{%
\'{E}cole Polytechnique F\'{e}d\'{e}rale de Lausanne (EPFL), Institute of Physics, CH-1015 Lausanne, Switzerland
}%
\orcid{0000-0001-7491-3660}
    \affiliation{
    Center for Quantum Science and Engineering, \'{E}cole Polytechnique F\'{e}d\'{e}rale de Lausanne (EPFL), CH-1015 Lausanne, Switzerland
    }
 \email{jannes.nys@epfl.ch}
    
\author{Giuseppe Carleo}
\affiliation{%
\'{E}cole Polytechnique F\'{e}d\'{e}rale de Lausanne (EPFL), Institute of Physics, CH-1015 Lausanne, Switzerland
}%
\orcid{0000-0002-8887-4356}
    \affiliation{
    Center for Quantum Science and Engineering, \'{E}cole Polytechnique F\'{e}d\'{e}rale de Lausanne (EPFL), CH-1015 Lausanne, Switzerland
    }

\begin{abstract}
    Through the introduction of auxiliary fermions, or an enlarged spin space, one can map local fermion Hamiltonians onto local spin Hamiltonians, at the expense of introducing a set of additional constraints. We present a variational Monte-Carlo framework to study fermionic systems through higher-dimensional (>1D) Jordan-Wigner transformations.  We provide exact solutions to the parity and Gauss-law constraints that are encountered in bosonization procedures. We study the $t$-$V$ model in 2D and demonstrate how both the ground state and the low-energy excitation spectra can be retrieved in combination with neural network quantum state ansatze.
\end{abstract}

\maketitle

\section{Introduction}
Studying the mapping between fermionic operators and bosonic operators (and vice versa) is interesting both from a theory perspective, as well as for computational studies. As examples of the former, the 1D transverse-field Ising model and the Kitaev honeycomb model can be diagonalized after reformulating the Hamiltonian in terms of fermionic degrees of freedom~\cite{kitaev2006anyons}. Furthermore, transforming fermionic Hamiltonians into a set of spin operators is necessary to compute properties of fermionic systems using digital quantum devices. Especially in the NISQ era~\cite{preskill2018quantum} of intermediate scale devices, it is highly advantageous to study efficient mappings that require fewer qubit resources.

The most natural and commonly used mapping between fermionic and spin degrees of freedom is the Jordan-Wigner transformation (JWT), which follows as a natural consequence
of the second quantization formalism of fermions~\cite{jordan1993paulische}. After we have chosen a fermion ordering in the second quantization formalism, the JWT maps each fermion operator $f_i^\dagger$ onto spin operators as follows:
% \footnote{We identify $\ket{0} = \ket{\uparrow}$ and $\ket{1} = \ket{\downarrow}$.} 
$f_i^\dagger \to  Q_i^-\otimes_{j<i} Z_j$, where $Q_i^- = (X_i-iY_i)/2$ and $(X_i, Y_i, Z_i)$ are Pauli matrices applied to spin $i$. The operator chain $S_i = \otimes_{j<i} Z_j$ is necessary to maintain the anti-commutation relations on the fermionic side using a set of Pauli matrix operators that themselves follow commutation relations, and are commonly referred to as Jordan-Wigner strings. Physical fermionic Hamiltonians that describe closed quantum systems consist of bilinear and quadratic terms in the creation/annihilation operators. Such Hamiltonians conserve the fermion parity $\mathcal{P}_f = (-1)^{N_f}$ (with $N_f$ the number of fermions), and are referred to as even parity operators. When the original fermionic operators are both spatially local and have even parity (i.e.\ conserve $\mathcal{P}_f$), the locality is trivially preserved in the resulting spin Hamiltonian in 1D, since the Jordan-Wigner strings $S_i$ of local fermionic operator pairs cancel each other. In higher dimensions (>1D), however, the chosen ordering of the fermions in the JWT becomes increasingly important. Local even-parity fermionic operators are no longer mapped onto a set of local products of Pauli matrices since the Jordan-Wigner strings $S_i$ of spatially local fermion operator pairs no longer cancel each other. When the dimensionality of the system increases, the Jordan-Wigner strings in the spin Hamiltonian become increasingly non-local and it therefore increasingly difficult to study these systems numerically~\cite{whitfield2016local}.

The mapping of fermion operators onto quantum spin operators is not unique. One can use this freedom in order to generalizations to fermion-spin mappings in higher dimensions with the main aim to maintain locality in the operators, and thereby reducing the size of the Jordan-Wigner strings. One of the first studies that derived higher-dimensional generalizations to the Jordan-Wigner transformation dates back to the work of Wosiek~\cite{wosiek1981local}. In this work, Wosiek described a mapping of fermions moving on a 2D and 3D square lattice onto a set of local generalized Euclidean Dirac matrices. Thereby, he found the need to impose additional constraints on the system in order to remove redundant and unphysical sectors of the new Hilbert space. The constraints generated in this bosonization procedure were studied numerically only recently in Ref.~\cite{bochniak2020constraints}.
Similar ideas were later explored by Bravyi and Kitaev~\cite{bravyi2002fermionic} in the simulation of fermionic systems through local qubit gates. As in Ref.~\cite{verstraete2005mapping}, they explored methods that increase the Hilbert space, while afterwards restricting the reachable quantum states to a physical sector of Hilbert space through a set of gauge conditions. Ball~\cite{ball2005fermions} demonstrated how these gauge conditions can be made local as well. 
Ball~\cite{ball2005fermions} and Verstraete-Cirac~\cite{verstraete2005mapping} both suggested the more explicit introduction of auxiliary fermionic modes to counteract the Jordan-Wigner strings. These auxiliary modes effectively store the parity nearby the interaction terms, which is otherwise captured by the Jordan-Wigner string~\cite{whitfield2016local}. The auxiliary Majorana fermions are subject to local interaction terms that commute with the physical Hamiltonian, which is necessary in order to keep the eigenspectrum of the original problem identifiable in the spectrum of the transformed Hamiltonian. 

In recent years, we have witnessed a renewed interest in methods for simulating fermionic systems through a set of local qubit gates~\cite{po2021symmetric, setia2018bravyi, setia2019superfast, chen2018exact, chen2022equivalence}. This more recent theoretical activity is again driven by two main motives. On one hand, local mappings are of practical interest in the implementation of quantum algorithms to simulate fermionic matter on increasingly available qubit-based digital quantum computers. On the other hand, a recent theoretical advance has been made in connecting bosonization in 2D and $Z_2$ lattice gauge theories with Chern–Simons-like Gauss laws~\cite{chen2018exact} (later generalized to three and arbitrary spatial dimensions in Ref.~\cite{chen2019bosonization, chen2020exact}). Various followup works have built on this connection to derive new fermion mappings in higher dimensions~\cite{li2021higher, po2021symmetric, bochniak2020bosonization, tantivasadakarn2020jordan}.
Compared to earlier methods (such as Refs.~\cite{ball2005fermions, verstraete2005mapping}) where JWT were carried out explicitly, recent techniques~\cite{chen2018exact, bochniak2020bosonization,bochniak2020constraints} take a different approach where one first defines bosonic operators from the fermionic ones, which can then be mapped directly onto quantum spin operators without the need to order the fermions. The equivalence of these bosonization procedures in 2D was proven by Chen~\etal\cite{chen2022equivalence}.

Despite this recent theoretical progress, the practical application of these techniques remains elusive, both in the context of classical computational methods, and as a basis for quantum algorithms. The main difficulty lies in the fact that the auxiliary degrees of freedom must satisfy stringent constraints in order to correctly represent fermionic degrees of freedom. Efficiently satisfying these constraints is a particularly important task especially in applications involving variational searches of many-body fermionic states. This is for example relevant for both classical variational methods based on spin/qubit degrees of freedom and for variational quantum algorithms tailored to qubits. In this work, we specifically focus on the variational simulation of fermionic systems on classical computers, with suitable many-body quantum states of spins degrees of freedom. Specifically, we demonstrate that we can factorize the wave function into an exact solution to the constraint, and a physical wave function that can be determined variationally. Furthermore, maintaining spatial locality in the transformation can allow us to map the fermionic Hamiltonian onto a spin Hamiltonian which features the same symmetries~\cite{li2021higher, po2021symmetric}, and therefore gives us access to the low-energy excitation spectrum. We then show that there is substantial variational freedom in parameterizing the resulting many-body state. In this context, we concentrate on on neural-network-based parameterizations of the many-body wave function, known as neural-network quantum states (NQS) ~\cite{Carleo17}. We show how this approach can be used to approximate the energy eigenstates of the fermionic system, using methods that are available to approximate quantum spin states~\cite{Carleo17, Choo18}.

\section{Bosonization}
\begin{figure}[tbh]
    \centering
    \includegraphics[width=0.45\textwidth, trim=50mm 40mm 30mm 100mm, clip]{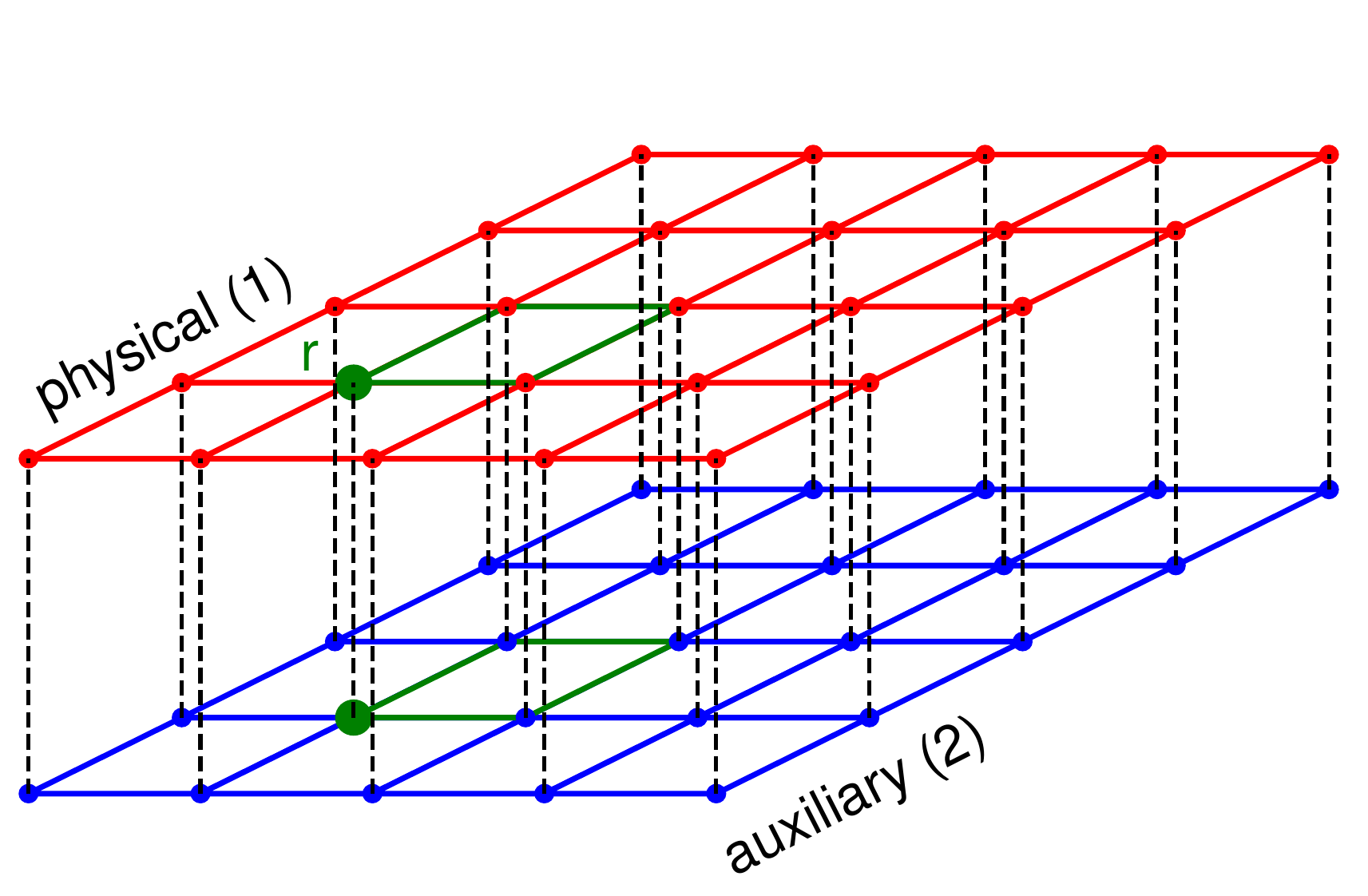}
    \caption{Illustration of the two sets of layers, containing a physical qubit layer and a set of auxiliary qubits. In green, we indicate the plaquette attached to position $\vec{r}$.}
    \label{fig:layers}
\end{figure}

We define an $L_x \times L_y$ 2D lattice with square cells and such that all edges point either along the lattice basis vectors $\vec{x}$ or $\vec{y}$. The resulting set of edges reads $\mathcal{E} = \{(\vec{r}, \vec{r}+\vec{x})| \vec{r} \in \mathcal{V}\} \cup \{(\vec{r}, \vec{r}+\vec{y})| \vec{r}\in \mathcal{V}\}$. Here, $\vec{x}$ and $\vec{y}$ represent the lattice vectors and $\vec{r}=(r_x, r_y)$, where $r_x\in \{0,..., L_x-1\}$ and $r_x = L_x$ maps onto $r_x = 0$ due to periodicity (similar for $r_y$).
We study the $t$-$V$ model (also called spinless Fermi-Hubbard model)
\begin{align}
\begin{split}
    H = &-t \sum_{(\vec{r}, \vec{r}') \in \mathcal{E}} \left[f^\dagger_{\vec{r}} f_{\vec{r}'} + f^\dagger_{\vec{r}'} f_{\vec{r}}\right] \\
    &+ V \sum_{\vec{r} \in \mathcal{V}} \sum_{\delta\vec{r} \in \{\vec{x}, \vec{y}\}} n_{\vec{r}} n_{\vec{r}+\delta\vec{r}} - \mu \sum_{\vec{r} \in \mathcal{V}} n_{\vec{r}} 
    \label{eq:hamiltonian_fermion}
\end{split}
\end{align}
where $f^\dagger_{\vec{r}}$ are fermionic operators, and we have introduced the usual number operator $n_{\vec{r}} = f^\dagger_{\vec{r}} f_{\vec{r}}$.
On each site, the physical Hilbert space is 2 dimensional $\{\ket{0},  f^\dagger_{\vec{r}}\ket{0}\}$, and the Hamiltonian in Eq.~\eqref{eq:hamiltonian_fermion} contains only even parity fermionic operators.
We carry out the bosonization procedure defined in Refs.~\cite{li2021higher, po2021symmetric}, since it keeps the symmetries of the Hamiltonian manifest. For the sake of completeness, we describe the procedure in Appendix~\ref{sec:bosonization}, and summarize the results here. Bosonizing the Hamiltonian in Eq.~\eqref{eq:hamiltonian_fermion} results in:
\begin{align}
\begin{split}
    H =& -\frac{t}{2} \sum_{\vec{r}\in \mathcal{V}}   (Y^{(1)}_{\vec{r}}Y^{(1)}_{\vec{r}+\vec{x}} + X^{(1)}_{\vec{r}}X^{(1)}_{\vec{r}+\vec{x}})Z^{(2)}_{\vec{r}+\vec{x}} \\ 
    &-\frac{t}{2} \sum_{\vec{r}\in \mathcal{V}} 
    (-X^{(1)}_{\vec{r}}Y^{(1)}_{\vec{r}+\vec{y}} + Y^{(1)}_{\vec{r}}X^{(1)}_{\vec{r}+\vec{y}}
    )Y^{(2)}_{\vec{r}} X^{(2)}_{\vec{r}+\vec{y}}  \\
    &+ \frac{V}{2} \sum_{\vec{r} \in \mathcal{V}} (1-2Z^{(1)}_{\vec{r}}) + \frac{V}{4} \sum_{\vec{r} \in \mathcal{V}} \sum_{\delta\vec{r} \in \{\vec{x}, \vec{y}\}} Z^{(1)}_{\vec{r}} Z^{(1)}_{\vec{r}+\delta\vec{r}}    \\
    &- \frac{\mu}{2} \sum_{\vec{r} \in \mathcal{V}} (1 - Z^{(1)}_{\vec{r}}) 
    \label{eq:spinless_ham_boson}
\end{split}
\end{align}
On the resulting square lattice, each site hosts a physical (1) and auxiliary qubit/spin (2), as demonstrated in Fig.~\ref{fig:layers}. As shown by Ref.~\cite{chen2022equivalence}, other bosonisation procedures are equivalent in 2D.

\subsection{Local constraints}
In order to reduce the number of degrees of freedom, the auxiliary system is subject to a Gauss-law constraint of the form 
\begin{align}
    G_{\vec{r}} &\overset{c}{=} 1 \qquad \forall \vec{r} \in \mathcal{V}\label{eq:gausslaw}
\end{align}
In our notation, constraints such as the ones in Eqs.~\eqref{eq:gausslaw} should be interpreted as constraints on the corresponding Hilbert space, i.e.\ $G_{\vec{r}} \ket{\Psi} \overset{c}{=} \ket{\Psi}$.
In terms of Pauli operators, $G_{\vec{r}}$ in Eq.~\eqref{eq:gausslaw} takes the form
\begin{align}
    G_\vec{r} &= [Z^{(1)} Y^{(2)} ]_{\vec{r}} [X^{(2)} ]_{\vec{r}+\vec{x}} [Y^{(2)} ]_{\vec{r}+\vec{x}+\vec{y}} [Z^{(1)} X^{(2)})]_{\vec{r}+\vec{y}}\label{eq:gaussG}
\end{align}
We can separate the physical (1) and auxiliary (2) system and rewrite Eq.~\eqref{eq:gausslaw} with the form in Eq.~\eqref{eq:gaussG} as
\begin{align}
Y^{(2)}_{\vec{r}} X^{(2)}_{\vec{r}+\vec{x}} Y^{(2)}_{\vec{r}+\vec{x}+\vec{y}} X^{(2)}_{\vec{r}+\vec{y}} \overset{c}{=} Z^{(1)}_{\vec{r}}Z^{(1)}_{\vec{r}+\vec{y}}\label{eq:gaussLR}
\end{align}
The operator on the left-hand side of this constraint
\begin{align}
    C_{\vec{r}} = Y^{(2)}_{\vec{r}} X^{(2)}_{\vec{r}+\vec{x}} Y^{(2)}_{\vec{r}+\vec{x}+\vec{y}} X^{(2)}_{\vec{r}+\vec{y}}\label{eq:gaussC}
\end{align}
also appears in Wen's plaquette model~\cite{wen2003quantum}, and hence, Eq.~\eqref{eq:gaussLR} can be interpreted as a dynamical version this model, due to the right-hand side which depends on the physical system. The constraint in Eq.~\eqref{eq:gaussLR} resembles a Chern-Simons Gauss law, or flux attachment~\cite{chen2018exact}. The constraint is diagonal in the physical system, and therefore, for each configuration of the physical system, the auxiliary system is in an eigenstate of the exactly solvable Wen's plaquette model (which is known to have robust topologically degenerate ground states), with different signs for the terms in the Hamiltonian. It is important to emphasize that the constraint in Eq.~\eqref{eq:gaussLR} is `kinematic', meaning it only depends on the lattice topology, not on the Hamiltonian under consideration. 

\subsection{Parity constraints}
We restrict the system to square even-by-even tori (i.e.\ $L = L_x = L_y$ is even) with $N=L^2$ sites. Imposing the boundary conditions in the fermionic system, we obtain the additional constraints introduced by non-contractable Wilson loops. After bosonization, we obtain the following spin operator identities that need to be satisfied
\begin{align}
     \oprod_{m=0}^{L_x-1} Z^{(1)}_{\vec{r}+m\vec{x}} Z^{(2)}_{\vec{r}+m\vec{x}} &\overset{c}{=} -1 \label{eq:pbc_constraint_x} \\      
     \oprod_{m=0}^{L_y-1} Z^{(2)}_{\vec{r}+m\vec{y}} &\overset{c}{=} -1 \label{eq:pbc_constraint_y}
\end{align}
Furthermore, we fix the number of fermions to be $N_f$, which is enforced through the constraint
\begin{align}
    \sum_{\vec{r}\in \mathcal{V}}\frac{1-Z^{(1)}_{\vec{r}}}{2} \overset{c}{=} N_f \label{eq:Nf_constraint}
\end{align}
We can now set the chemical potential $\mu = 0$, since it only adds a constant energy.

\section{Variational Monte Carlo approach}\label{sec:vmc}
The goal of this paper is to obtain variational solutions to the bosonized fermionic system. Hereby, we rely on Variational Monte Carlo (VMC). We briefly recap the concepts of VMC within our notation. For a more elaborate and pedagogical introduction, we refer to e.g.\ Ref.~\cite{vicentini2021netket}. After a short recap, we will introduce our novel approaches to solving the gauge constraints described above.

%\subsection{Introduction to VMC}
We will study the system in Eq.~\eqref{eq:hamiltonian_fermion} using the bosonized Hamiltonian in Eq.~\eqref{eq:spinless_ham_boson}.  Our aim is to obtain the ground and low-lying excited states $\ket{\Psi}$ using the decomposition
\begin{align}
    \ket{\Psi} = \sum_{\vec{\sigma}} \Psi_\theta(\vec{\sigma}) \ket{\vec{\sigma}}\label{eq:decomp}
\end{align}
In this notation $\ket{\vec{\sigma}}$ represent basis states in the $S_z$ basis. The systems consists of a physical system $\vec{\sigma}^{(1)} = (\sigma_{\vec{r}_{1}}^{(1)}, ..., \sigma_{\vec{r}_N}^{(1)})$ and auxiliary system $\vec{\sigma}^{(2)} = (\sigma_{\vec{r}_{1}}^{(2)}, ..., \sigma_{\vec{r}_N}^{(2)})$, such that $\vec{\sigma} = (\vec{\sigma}^{(1)}, \vec{\sigma}^{(2)})$. We will take $\sigma_{\vec{r}} \in \{-1, +1\}$ for simplicity.
The probability amplitudes in Eq.~\eqref{eq:decomp} are given by a parametrized function $\Psi_\theta$ with parameters $\theta$. These parameters are determined by minimizing the variational energy
\begin{align}
\begin{split}
    \braket{E}_\theta &= \frac{\braket{\Psi_\theta|H|\Psi_\theta}}{\braket{\Psi_\theta|\Psi_\theta}}\\
    &= \left< \sum_{\vec{\sigma}'} \braket{\vec{\sigma}|H|\vec{\sigma}'} \frac{\Psi_\theta(\vec{\sigma}')}{\Psi_\theta(\vec{\sigma})}\right>_{\vec{\sigma}\sim \abs{\Psi_\theta}^2}\label{eq:Eloc}
\end{split}
\end{align}
We will evaluate the expectation value $\left<\phantom{H}\right>_{{\vec{\sigma}\sim \abs{\Psi_\theta}^2}}$ in Eq.~\eqref{eq:Eloc} by sampling spin configurations using Markov-Chain Monte Carlo (MCMC). However, the constraints in Eqs.~\eqref{eq:gausslaw}, and \eqref{eq:pbc_constraint_x}--\eqref{eq:Nf_constraint} restrict the allowed Hilbert space, and hence we must enforce these restrictions either by imposing constraints on $\Psi_\theta$, and/or by designing Markov-Chain samplers that only generate samples within the allowed Hilbert space.

\subsection{Solving constraints in VMC}
We now describe our novel approach to satisfy all constraints. To the best of our knowledge, there have so far been no attempts to solve local fermion-to-qubit transformed Hamiltonians in the VMC framework. The constraints in Eq.~\eqref{eq:pbc_constraint_x}, ~\eqref{eq:pbc_constraint_y} and ~\eqref{eq:Nf_constraint} can be fulfilled exactly through a suitable sampling procedure since the constraints are diagonal in the $S_z$ basis. More specifically, MCMC samples can be generated through a set of sample updates based on the (free-fermion) Hamiltonian. Hence, given a sample, we can make the following Markov transitions
\begin{enumerate}
    \item (if $\sigma^{(1)}_{\vec{r}} \neq \sigma^{(1)}_{\vec{r}+\vec{x}}$): flip two neighboring qubits of system (1) along the $x$-axis: $X^{(1)}_{\vec{r}} X^{(1)}_{\vec{r}+\vec{x}}$, or
    \item (if $\sigma^{(1)}_{\vec{r}} \neq \sigma^{(1)}_{\vec{r}+\vec{y}}$): flip four qubits (2 physical, 2 auxiliary) on two neighboring sites along the $y$-axis: $X^{(1)}_{\vec{r}} X^{(1)}_{\vec{r}+\vec{y}} X^{(2)}_{\vec{r}} X^{(2)}_{\vec{r}+\vec{y}}$
\end{enumerate}
It is also straightforward to generate initial random samples that fulfill the parity and number constraints, which are necessary to initiate the Markov chains. First, a physical system (1) fulfilling the $N_f$ constraint in Eq.~\eqref{eq:Nf_constraint} can be constructed. Next, we generate $(L_x-1) \times (L_y-1)$ random auxiliary qubit states, and infer the remaining auxiliary qubits by imposing the periodicity constraints in Eqs.~\eqref{eq:pbc_constraint_x} and \eqref{eq:pbc_constraint_y}. The resulting scheme is summarized in Algorithm~\ref{alg:randomstates}. Alternatively, these parity constraints can be captured by a Restricted Boltzmann Machine (RBM) quantum state by introducing a hidden neuron with specific weight function~\cite{lu2019efficient}.\\

The Gauss law in Eq.~\eqref{eq:gausslaw} cannot be satisfied through a sampling procedure alone, since the operators are not diagonal in the computational basis, and they therefore impose stringent constraints on the wave function itself.

Within the basis that fulfills the constraints in Eq.~\eqref{eq:pbc_constraint_x} and \eqref{eq:pbc_constraint_y} one can obtain the eigenspectrum of the original Hamiltonian through eigendecomposition of $P_G^{-1} H  P_G$, where $P_G$ is the projection operator to the Gaussian constraint in Eq.~\eqref{eq:gausslaw}
\begin{align}
    P_G = \oprod_{\vec{r} \in \mathcal{V}} \frac{1 + G_\vec{r}}{\sqrt{2}}
\end{align}
While the representation of the $P_G$ operator in the $S_z$ basis is block diagonal, it is not sparse, since the size of each block scales exponentially with the system size. Furthermore, since constraint Eq.~\eqref{eq:gausslaw} must be satisfied for all $\vec{r} \in \mathcal{V}$, the projection generates non-local effects in the auxiliary system, even though the individual constraints are local.
Therefore, we must find an analytical form for the probability amplitude of a general quantum state that lies within the manifold of the Gauss law. Another approach is to implement the constraint in the Hamiltonian by adding the terms
\begin{align}
    H_c = - K \sum_{\vec{r}\in \mathcal{V}} G_{\vec{r}}
\end{align}
When the coupling $K$ is taken to be sufficiently large, $G_{\vec{r}} = 1$ can in principle be satisfied by minimizing the total energy of the augmented Hamiltonian $H' = H + H_c$ (this procedure is also suggested in Ref.~\cite{verstraete2005mapping}). In practice, however, we find that a soft constraint does not result in quantum states lying on the manifold dictated by the Gauss law, and therefore the spectrum does not reliably represent the physics of the fermionic system.

As mentioned, exactly respecting Eq.~\eqref{eq:gausslaw} is essential in order to restrict our solution to the physical Hilbert space representing fermionic degrees of freedom. Since the r.h.s.\ operator in Eq.~\eqref{eq:gaussLR} is diagonal in the physical states, each Gauss constraint can be regarded as a dynamical constraint on the auxiliary system. It is important to obtain variational ansatzes which obey the Gauss law constraints by construction. Our (non-symmetrized) variational ansatz in general assumes a factorized form
\begin{align}
    \braket{\vec{\sigma}|\Psi_\theta} = \Psi_\theta(\vec{\sigma}) = \xi(\vec{\sigma}) \Phi_\theta (\vec{\sigma}^{(1)})\label{eq:generalfactorized}
\end{align}
where $\theta$ represent a set of variational parameters, and $\xi(\vec{\sigma})$ is purely a sign factor with $\abs{\xi} = 1$. In this representation, we choose $\xi(\vec{\sigma})$ to be the only factor with an explicit dependence on the auxiliary system. Notice that there are no constraints on $\Phi_\theta$ with respect to anti-symmetry, since this property is entirely covered by the $\xi$ parity factor. Hence, $\xi$ must be chosen in such a way that Eq.~\eqref{eq:gausslaw} is exactly fulfilled. However, we point out that many solutions are feasible due to the freedom of extracting additional sign structure from $\Phi_\theta$, and absorbing it into $\xi$. Indeed, since $G_\vec{r}$ is diagonal in the physical system (1), we obtain independent Gauss-law constraints for each physical configuration $\vec{\sigma}^{(1)}$. 
After defining the sign-generating function $\xi$, the sign structure of $\Phi(\vec{\sigma}^{(1)})$ is determined by the local spin Hamiltonian in Eq.~\eqref{eq:spinless_ham_boson}. Furthermore, the main challenge is to find an exact solution that satisfies \emph{all} constraints, including the parity constraints in Eqs.~\eqref{eq:pbc_constraint_x}-\eqref{eq:pbc_constraint_y}.

\subsection{Canonical sample reduction}\label{sec:solution_reduce}
As a first solution to the Gauss law constraint, we define a consistent approach which uses a repeated application of $G_{\vec{r}}$ to a quantum state to transform each auxiliary configuration $\vec{\sigma}^{(2)}$ to a reduced auxiliary sample $\vec{\alpha}^{(2)}$. One can verify that the following ansatz is an eigenstate of the constraint operators $G_\vec{r}$:
\begin{align}
    \Psi_\theta (\vec{\sigma}) = &\phantom{\times} \braket{\vec{\sigma}^{(2)}|C_{\vec{r}_1}^{m_1}...C_{\vec{r}_N}^{m_{\vec{r}_N}}|\vec{\alpha}^{(2)}} \nonumber\\
    &\times (\sigma^{(1)}_{\vec{r}_1} \sigma^{(1)}_{\vec{r}_1+\vec{y}})^{m_1}...(\sigma^{(1)}_{\vec{r}_N} \sigma^{(1)}_{\vec{r}_N+\vec{y}})^{m_N} \nonumber \\
    &\times \Phi_\theta (\vec{\sigma}^{(1)})
\end{align}
Here, $m_i \in \{0,1\}$ is determined by the iterative procedure followed to obtain the reduced sample. Therefore, we define an ordering for the lattice sites $(\vec{r}_1$, ..., $\vec{r}_N)$ and iteratively set $m_i = 1$ if the auxiliary qubit at position $\vec{r}_i$ is in the $\ket{1}$ state, and $m_i = 0$ otherwise. When $m_i=1$, we apply operator $C_{\vec{r}_i}$ to the auxiliary plaquette attached to $\vec{r}_i$ and continue with the next site in the sequence $\vec{r}_{i+1}$. Algorithm~\ref{alg:canred} describes the sequential steps to obtain either directly $\xi(\vec{\sigma})$, or $(m_1,...,m_N)$ to evaluate Eq.~\eqref{eq:generalfactorized}. Also note that to each physical system configuration $\vec{\sigma}^{(1)}$, there corresponds only a single $\vec{\alpha}^{(2)}$. The time complexity of this approach to satisfy the Gauss constraint is $\order{N}$ for each sample $\vec{\sigma}$.

\subsection{Doubly canonical sample reduction}\label{sec:doubly_canonical}
In the abovementioned solution, the symmetry properties of $\Phi(\vec{\sigma}^{(1)})$ are elusive. Alternatively, we can choose to optimize the sign structure of $\Phi$ by incorporating knowledge from the relevant symmetry group.
Therefore, the above-mentioned procedure can also be carried out on the reduced samples obtained by listing the samples obtained by applying all symmetry elements $g(\vec{\sigma})$ in the symmetry group $g \in G$. We then take the sample $\vec{\sigma}_{\textrm{red}}$ with the smallest lexicographic encoding of the physical system and apply the above-mentioned sequence of reductions on $\vec{\sigma}_{\textrm{red}}$ to obtain $\vec{\alpha}^{(2)}$. The representation of symmetry operations will be discussed in more detail in Section~\ref{sec:intro_symmetries}.

\subsection{Vacuum reduction}\label{sec:vacuum_reduction}
A final method approaches the problem differently by defining the correct sign structure $\xi$ only explicitly on the vacuum state $\ket{\vec{0}^{(1)}\vec{\sigma}^{(2)}}$, and inferring the signs of other configurations by consistently relating them to the vacuum. The latter is defined in terms of spin states as
\begin{align}
\ket{\Omega_q} = \sum_{\vec{\sigma}^{(2)}} c_{\vec{\sigma}^{(2)}}\ket{\vec{0}^{(1)}, \vec{\sigma}^{(2)}}\label{eq:vacuum}
\end{align}
where $c_{\vec{\sigma}^{(2)}} = \xi(\vec{0}^{(1)},\vec{\sigma}^{(2)}) = \pm 1$ is obtained via the approach in Section~\ref{sec:solution_reduce}, and thus satisfies Eq.~\eqref{eq:gausslaw}. In the vacuum, the constraint in Eq.~\eqref{eq:gaussG} reduces to $C_{\vec{r}}=1$.

In order to obtain $\xi(\vec{\sigma})$ from the definition of $\ket{\Omega_q}$, we take the set of occupation numbers $(n_1, ..., n_N)$ corresponding to $\vec{\sigma}^{(1)}$ using $n_i = (1-\sigma_i^{(1)})/2$), and rewrite these in terms of fermion operators in the usual way $\ket{n_1, ..., n_N} = f_{\vec{r}_{N_f}}^\dagger ... f_{\vec{r}_1}^\dagger \ket{\Omega_f}$. The fermionic operator $f_{\vec{r}_{N_f}}^\dagger ... f_{\vec{r}_1}^\dagger$ can be transformed into a set of spin operator through the same bosonisation procedures that led to the spin Hamiltonian in Eq.~\eqref{eq:spinless_ham_boson}.
Our ansatz is now inspired by the above-mentioned mapping, and hence we assume
\begin{align}
    \xi(\vec{\sigma}) = \braket{\vec{\sigma}^{(1)}\vec{\sigma}^{(2)}|\mathcal{B}\left[f_{\vec{r}_{N_f}}^\dagger ... f_{\vec{r}_1}^\dagger\right]| \Omega_q} ~\label{eq:xi3}
\end{align}
where $\mathcal{B}$ represents the bosonisation procedure from Ref.~\cite{po2021symmetric}, using the mappings in Eq.~\eqref{eq:jw}. Notice that to consistently bosonize non-local fermionic operators $f_{\vec{r}_{N_f}}^\dagger ... f_{\vec{r}_1}^\dagger$ in Eq.~\eqref{eq:xi3}, this procedure requires an ordering of the sites, similarly to JWT.
The above-mentioned solution also forms a solution to tackle the standard JWT Hamiltonian. The main conceptual difference that in the current formalism, the ansatz is varied to optimize a \emph{local} Hamiltonian with \emph{manifest symmetries}. However, as pointed out in Ref.~\cite{bochniak2020bosonization}, despite the fact that the constraints in Eq.~\eqref{eq:gausslaw} are local, they can introduce long-range correlations when they are satisfied for all plaquettes (which shows up in the sign structure).
It is important to point out that the variational part of our ansatz $\Phi_\theta (\vec{\sigma}^{(1)})$ is a function of the $\vec{\sigma}^{(1)}$, which is equivalent to an occupation configuration. 

The resulting ansatz exactly fulfills the Gauss law in Eq.~\eqref{eq:gausslaw}, and hence, the physical part of the wave function $\Phi(\vec{\sigma}^{(1)})$ is constraint free, meaning it does not require anti-symmetrization and therefore our method is determinant free. Hence, we may use a universal function approximator such as a Neural Network, to represent $\Phi_\theta(\vec{\sigma}^{(1)})$. In this work, we adopt a simple Restricted Boltzmann Machine (RBM) ansatz with complex weights and $N$ hidden spins (thus with a hidden-spin density of $\alpha = 1$) ~\cite{Carleo17}.

\begin{figure*}[tb!]
    \centering    
    \includegraphics[width=0.33\textwidth]{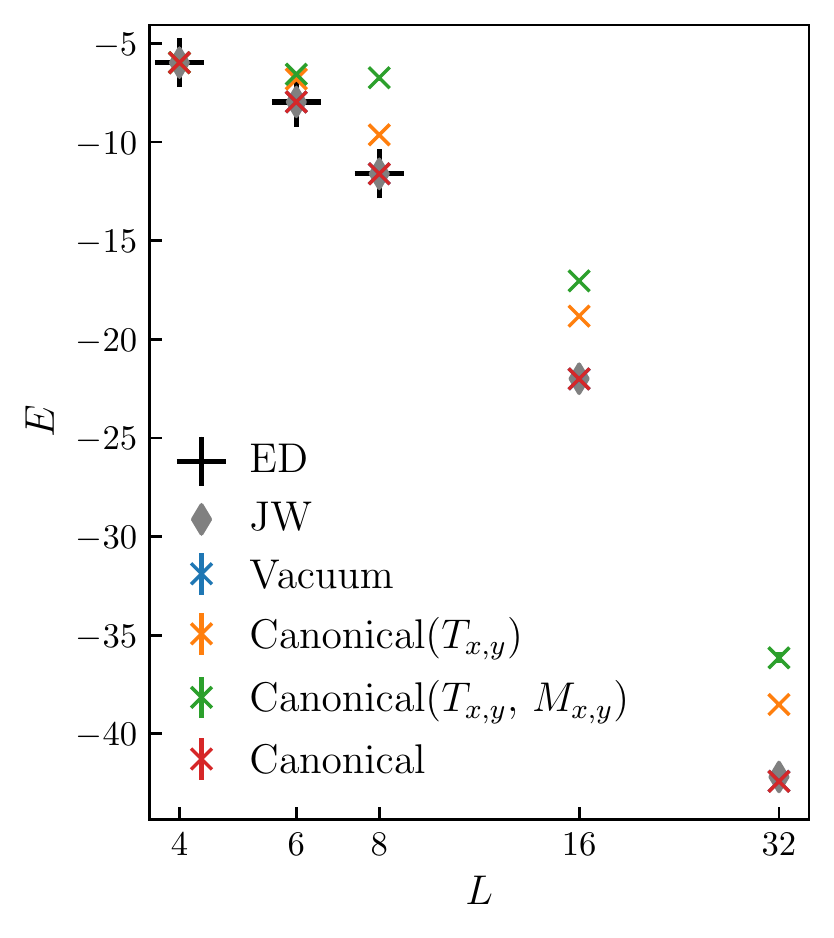}%
    \includegraphics[width=0.33\textwidth]{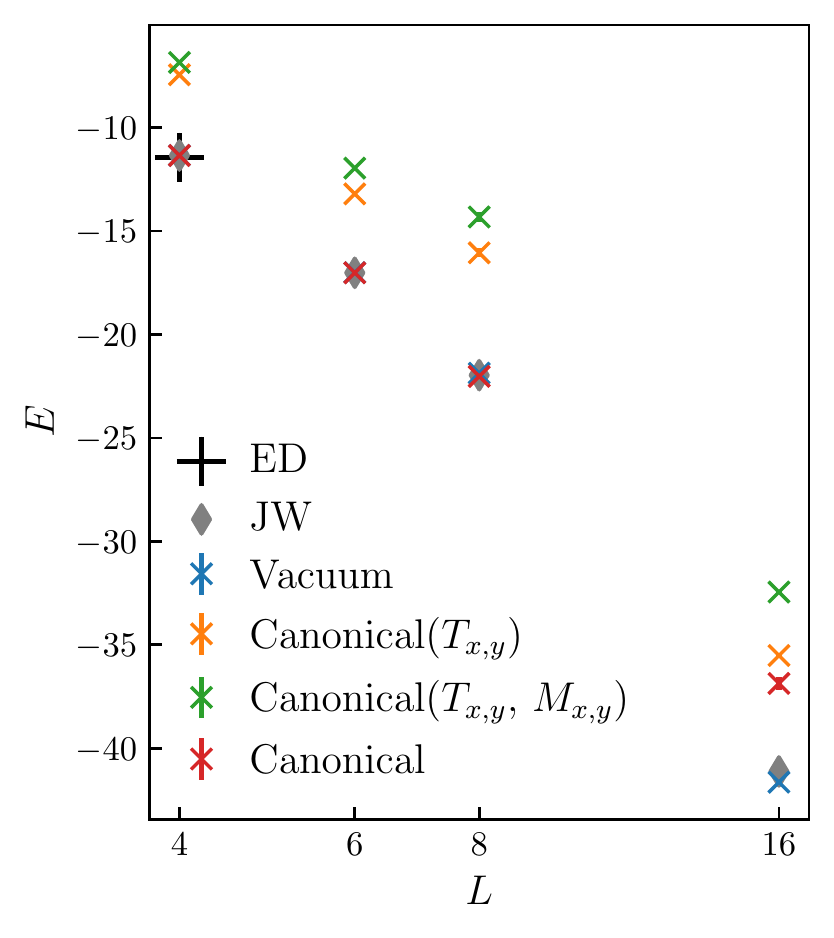}
    \caption{Ground state energies of models using the different strategies to respect the Gauss law constraint. The results are presented for lattices of (a) size $2 \times L$ with interaction strength $V/t=0.01$ and for (b) $4 \times L$  with $V/t=0.1$. Black crosses indicate the ground state obtained through Exact Diagonalization (ED). We separate the canonical sampling reduction from Section~\ref{sec:solution_reduce} (`canonical'), the doubly canonical reduction from Section~\ref{sec:doubly_canonical} (shown for reductions based on translation and translation+reflection symmetry), and the vacuum reduction from Section~\ref{sec:vacuum_reduction}. The statistical error bars are often too small to be visible.\label{fig:stretch1}}
\end{figure*}

\section{Quantum state symmetries}\label{sec:intro_symmetries}
It is expected that imposing symmetries will result in more reliable and accurate predictions of the ground state. We therefore turn to the representation of symmetry transformations within the bosonisation framework.
A representation $T_g$ of an element $g$ of a symmetry group $G$, can be decomposed into two components: a `bare' transformation $T^b$, and an auxiliary-mode transformation $V_{T}$, such that $U_{T} = V_T T^b$. The $V_T$ are tensor products of local single-site operators: $V_{T} = \oprod_{\vec{r} \in \mathcal{V}} V_{T, \vec{r}}$. These effectively replace the (non-local) parity factor one would encounter in the Jordan-Wigner transformation. Once we are able to determine $V_{T, \vec{r}}$ for all elements in the $C_{4v}$ group, we can obtain a symmetric quantum state ansatz (lying withing a chosen irrep $I$ of the group), using
\begin{align}
    \Psi^I(\vec{\sigma}) &= \frac{1}{\abs{G}}\sum_{g \in G} \chi^I_g \braket{\vec{\sigma} | U_{T_g} | g(\vec{\sigma})} \Psi(g(\vec{\sigma}))
\end{align}
where $\chi_g$ represents the character of the chosen irrep. 
For translations, we have
\begin{align}
	\Psi^{\vec{K}}(\vec{\sigma}) &= \frac{1}{\abs{G}}\sum_{g \in G} e^{i \vec{K}.\vec{R}_g } \Psi(g(\vec{\sigma}))
\end{align}
More details on symmetry operations and the $V_{T, \vec{r}}$ for translation, rotation and reflection symmetry are deferred to Appendix~\ref{sec:symmetry}.
Notice that the factor $\xi(g(\vec{\sigma}))$ inside $\Psi$ is the alternative for a factor which determines the parity of the $T_g$ operator on $\vec{\sigma}$. Hence, no sorting is required, and the parity is determined through local operators, since parity information is captured by the auxiliary modes.  The above-mentioned methods can therefore be carried out in $\order{N}$ time, contrary to other determinant-free methods Ref.~\cite{inui2021determinant}. The latter would also be necessary when carrying out 1D Jordan-Wigner procedures, in which one must compute the parity of the translation operator in terms of fermion operators.

\section{Results}

\begin{figure*}[htb]
    \centering
    \includegraphics[width=0.33\textwidth]{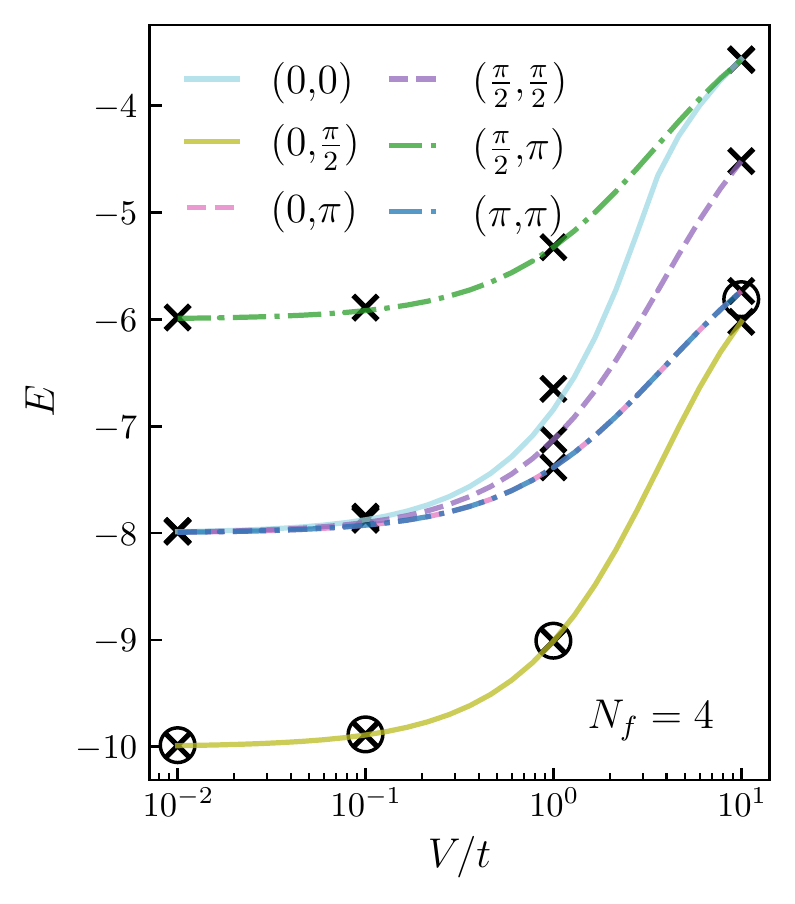}%
    \includegraphics[width=0.33\textwidth]{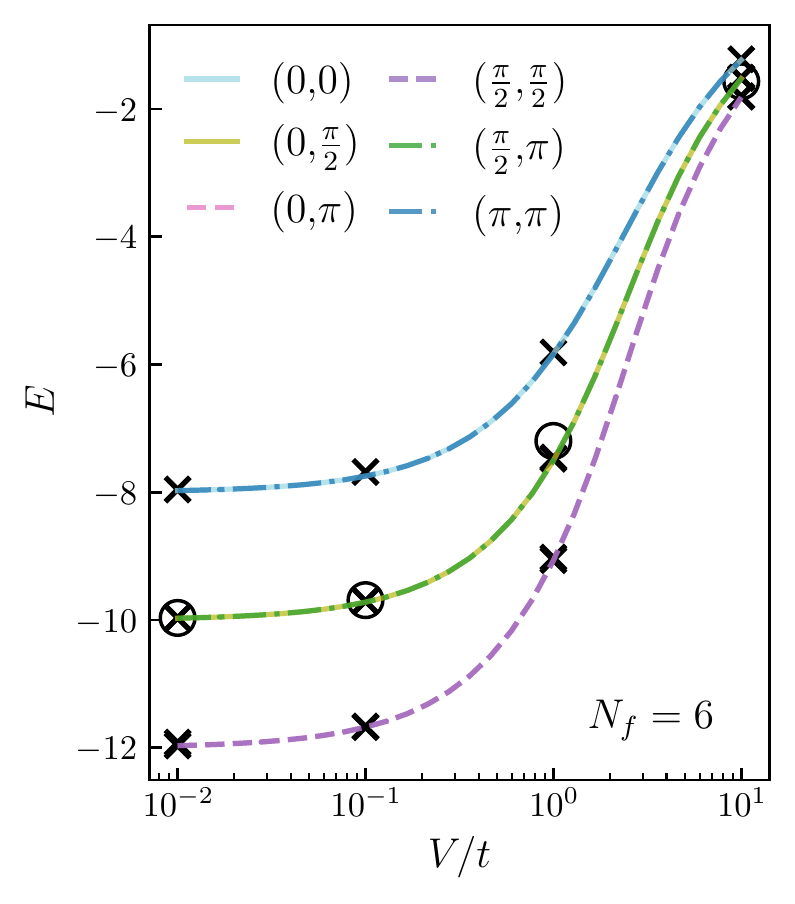}%
    \includegraphics[width=0.33\textwidth]{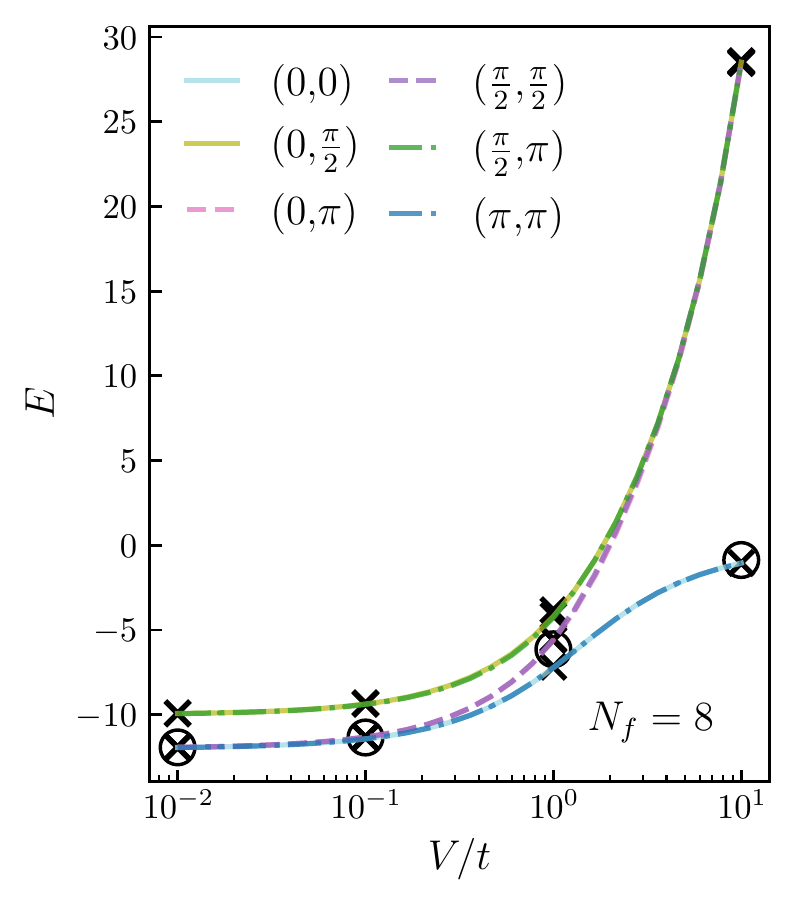}
    \caption{ Lowest energy levels in the relevant translation symmetry irreps (indicated with momentum $\vec{k}$) at different fixed number of fermions $N_f \in \{4,6,8\}$ at interaction potential $V/t\in \{10^{-2},10^{-1},1,10\}$ and a lattice of size $4 \times 4$. Colored lines are results from ED. Crosses are results of VMC optimization within the given irrep, and the circles represents an optimization without symmetry restrictions.}
    \label{fig:symmetry_results}
\end{figure*}

We carry out numerical studies of the bosonization procedures by relying on the above-mentioned solutions to all constraints. Hereby, we first investigate their ability to represent the ground state and the effect of different choices of $\xi(\vec{\sigma})$ on the complexity of learning $\Phi_\theta(\vec{\sigma}^{(1)})$. Through these studies, we mainly probe the sensitivity of numerical and variational approaches to the dynamical character of the Gauss law constraint in Eq.~\eqref{eq:gausslaw}, which was interpreted as a dynamical Wen plaquette model. Consider first an unsymmetrized variational ansatz. In Fig.~\ref{fig:stretch1}, we show how the different methods to satisfy the Gauss law perform for a  $2\times L$ lattices. We compare the results to the ones obtained through a Jordan-Wigner transformation mapping the two dimensional lattice onto a one-dimensional one through "snaking" along the $L$ direction. By increasing $L$, we increase the degree of non-locality in the Jordan-Wigner-transformed Hamiltonian, where Jordan-Wigner strings appear explicitly in the kinetic terms. Indeed, hopping operators transform under JWT as $f_\vec{r}^\dagger f_{\vec{r}+\vec{y}} \to Q_{\vec{r}}^- \left(\oprod_{\vec{r}\in P_{i\to j}} Z_{\vec{r}}\right)Q_{\vec{r}+\vec{y}}^+ $, where $P_{i\to j}$ represents all sites on the path along the snake direction connecting site $\vec{r}$ and $\vec{r}+\vec{y}$. In the extreme case of $\vec{r} = (0,0)$, the length of this path is $2L-2$.
Although the canonical and doubly canonical methods are valid solutions to the Gauss law constraints, the results demonstrate that the corresponding physical wave function factor $\Phi_\theta(\vec{\sigma}^{(1)})$ are challenging to optimize. The underlying reason is that the $\xi$ factors impose a non-trivial sign structure on $\Phi_\theta$. Although the doubly canonical method should simplify the sign structure to be learned for samples that are connected though the considered symmetry operations, the ground state does not necessarily correspond to trivial irrep characters. Furthermore, the variational wave function $\Phi_\theta$ must still capture the signs of samples that are not connected through symmetry operators, an effect that is reflected in the inaccurate ground-state representations with this method.
The vacuum reduction, however, only uses the non-local reduction method in Section~\ref{sec:solution_reduce} on the vacuum state (by solving the corresponding Wen plaquette model~\cite{wen2003quantum}). This results in energies closely matching the Exact Diagonalization (ED) results, and closely match those obtained with the snaked Jordan-Wigner approach. Similar conclusions can be drawn from Fig.~\ref{fig:stretch1}b, where we increased the short dimension of the lattice, as well as the interaction strength. A difference between the canonical and vacuum method becomes increasingly visible for the $4 \times L$ lattice at large $L$, where the vacuum reduction method tends to result in lower energies. Notice that the canonical and vacuum methods result in similar energies for $2\times L$ lattices since the canonicalization procedure operates on a single chain of plaquettes. On the other hand, for  $4\times L$ lattices, we follow a snaking procedure to canonicalize the plaquettes.

Next, we consider a $4\times 4$ system, and investigate the effect of embedding symmetries in the variational ansatz. We focus on  the vacuum reduction method from Section~\ref{sec:vacuum_reduction}, which generated superior results in the previous experiment. We compare the results to the ones obtainable through ED in Fig.~\ref{fig:symmetry_results}. Note that the Hamiltonian is diagonalized on the manifold determined by the total projector $P_G  P_x  P_y  P_{N_f}$, where the projectors $P_x$, $P_y$, $P_{N_f}$ enforce the parity constraints in Eqs.~\eqref{eq:pbc_constraint_x},~\eqref{eq:pbc_constraint_y} and~\eqref{eq:Nf_constraint} respectively. The unsymmetrized ansatz generates accurate ground-state energies, except for $N_f = 6$.
In the latter case, the ground state has a challenging sign structure, since the corresponding translation-symmetry sectors are $(\tfrac{\pi}{2}, \tfrac{\pi}{2})$ and $(0, \pi)$. On the other hand, $N_f = 4$ has an isolated ground state, and the ground-state for $N_f=8$ contains the simpler $(0,0)$ sector. Therefore, the ground states of $N_f = 4, 8$ are more accessible during the VMC optimization. Hence, while the variational ansatz can accurately represent the ground state, and is only hindered by the optimization procedure. Symmetrizing the ansatz removes the local minimum encountered in the optimization, and therefore returns significantly better ground state estimates. When the energy gap is small, the unsymmetrized ansatz does not accurately represent the ground state. The latter is not necessarily due to a limited representational power, but is directly related to the optimization procedure via Stochastic Reconfiguration (SR), which heavily depends on the size of the gap. Symmetrizing the ansatz and restricting to a given irrep opens the gap and therefore generates more accurate results.

\section{Conclusions}
We showed that bosonization procedures which allow one to transform local fermionic operators into local qubit operators result in Gauss law constraints that must be fulfilled in order to reduce the Hilbert space. Fulfilling all constraints at the same time is challenging, and we introduced multiple approaches that solve these constraints exactly, without the need to apply projection operators. We applied our method to the $t$-$V$ model on a square lattice in 2D. We argued that constraints related to parity and boundary conditions can be fulfilled through the sampling procedure. Our experiments demonstrate that satisfying the Gauss constraints directly affects our ability to reliably represent the ground state of a given Hamiltonian. We found that imposing a sign structure on the vacuum state simplifies this challenge. However, the gauge constraint introduces non-local effects, as would also be expected from a standard Jordan-Wigner transformation. By following the bosonization procedure from Ref.~\cite{po2021symmetric}, we keep the symmetries of the fermionic system manifest, which allows us to restrict the variational ansatz to a chosen symmetry irrep using techniques that are commonly used to study quantum spin systems. We demonstrate that this allows us to study the low-energy spectrum using neural-network quantum state ansatze. Since the Gauss constraint depends only on the lattice topology, our approach can in the future directly be applied to other Hamiltonians on square lattices. We foresee extensions of our work to other lattices and higher dimensions, as well as studies of the bosonic systems that are equivalent to fermionic systems.

\section*{Acknowledgement}
This work was supported by Microsoft Research.

\bibliographystyle{quantum}
\bibliography{biblio}% Produces the bibliography via BibTeX.

\begin{thebibliography}{10}

\bibitem{kitaev2006anyons}
Alexei Kitaev.
\newblock ``Anyons in an exactly solved model and beyond''.
\newblock
  \href{https://dx.doi.org/https://doi.org/10.1016/j.aop.2005.10.005}{Annals of
  Physics {\bf 321}, 2--111}~(2006).

\bibitem{preskill2018quantum}
John Preskill.
\newblock ``Quantum computing in the {NISQ} era and beyond''.
\newblock
  \href{https://dx.doi.org/https://doi.org/10.22331/q-2018-08-06-79}{Quantum
  {\bf 2}, 79}~(2018).

\bibitem{jordan1993paulische}
Pascual Jordan and Eugene~Paul Wigner.
\newblock ``{\"U}ber das paulische {\"a}quivalenzverbot''.
\newblock In The Collected Works of Eugene Paul Wigner.
\newblock \href{https://dx.doi.org/https://doi.org/10.1007/BF01331938}{Pages
  109--129}.
\newblock Springer~(1993).

\bibitem{whitfield2016local}
James~D Whitfield, Vojt{\v{e}}ch Havl{\'\i}{\v{c}}ek, and Matthias Troyer.
\newblock ``Local spin operators for fermion simulations''.
\newblock \href{https://dx.doi.org/10.1103/PhysRevA.94.030301}{Physical Review
  A {\bf 94}, 030301}~(2016).

\bibitem{wosiek1981local}
Jacek Wosiek.
\newblock ``A local representation for fermions on a lattice''.
\newblock Technical report.
\newblock Univ., Physics Department~(1981).
\newblock
  url:~\href{https://inspirehep.net/literature/169185}{inspirehep.net/literature/169185}.

\bibitem{bochniak2020constraints}
Arkadiusz Bochniak, B{\l}a{\.z}ej Ruba, Jacek Wosiek, and Adam Wyrzykowski.
\newblock ``Constraints of kinematic bosonization in two and higher
  dimensions''.
\newblock \href{https://dx.doi.org/10.1103/PhysRevD.102.114502}{Physical Review
  D {\bf 102}, 114502}~(2020).

\bibitem{bravyi2002fermionic}
Sergey~B Bravyi and Alexei~Yu Kitaev.
\newblock ``Fermionic quantum computation''.
\newblock
  \href{https://dx.doi.org/https://doi.org/10.1006/aphy.2002.6254}{Annals of
  Physics {\bf 298}, 210--226}~(2002).

\bibitem{verstraete2005mapping}
Frank Verstraete and J~Ignacio Cirac.
\newblock ``Mapping local hamiltonians of fermions to local hamiltonians of
  spins''.
\newblock
  \href{https://dx.doi.org/https://doi.org/10.1088/1742-5468/2005/09/P09012}{Journal
  of Statistical Mechanics: Theory and Experiment {\bf 2005}, P09012}~(2005).

\bibitem{ball2005fermions}
RC~Ball.
\newblock ``Fermions without fermion fields''.
\newblock \href{https://dx.doi.org/10.1103/PhysRevLett.95.176407}{Physical
  review letters {\bf 95}, 176407}~(2005).

\bibitem{po2021symmetric}
Hoi~Chun Po.
\newblock ``Symmetric {J}ordan-{W}igner transformation in higher
  dimensions''~(2021).

\bibitem{setia2018bravyi}
Kanav Setia and James~D Whitfield.
\newblock ``Bravyi-kitaev superfast simulation of electronic structure on a
  quantum computer''.
\newblock \href{https://dx.doi.org/https://doi.org/10.1063/1.5019371}{The
  Journal of chemical physics {\bf 148}, 164104}~(2018).

\bibitem{setia2019superfast}
Kanav Setia, Sergey Bravyi, Antonio Mezzacapo, and James~D Whitfield.
\newblock ``Superfast encodings for fermionic quantum simulation''.
\newblock \href{https://dx.doi.org/10.1103/PhysRevResearch.1.033033}{Physical
  Review Research {\bf 1}, 033033}~(2019).

\bibitem{chen2018exact}
Yu-An Chen, Anton Kapustin, and Đorđe Radičević.
\newblock ``Exact bosonization in two spatial dimensions and a new class of
  lattice gauge theories''.
\newblock
  \href{https://dx.doi.org/https://doi.org/10.1016/j.aop.2018.03.024}{Annals of
  Physics {\bf 393}, 234--253}~(2018).

\bibitem{chen2022equivalence}
Yu-An Chen and Yijia Xu.
\newblock ``Equivalence between fermion-to-qubit mappings in two spatial
  dimensions''~(2022).

\bibitem{chen2019bosonization}
Yu-An Chen and Anton Kapustin.
\newblock ``Bosonization in three spatial dimensions and a 2-form gauge
  theory''.
\newblock \href{https://dx.doi.org/10.1103/PhysRevB.100.245127}{Physical Review
  B {\bf 100}, 245127}~(2019).

\bibitem{chen2020exact}
Yu-An Chen.
\newblock ``Exact bosonization in arbitrary dimensions''.
\newblock \href{https://dx.doi.org/10.1103/PhysRevResearch.2.033527}{Physical
  Review Research {\bf 2}, 033527}~(2020).

\bibitem{li2021higher}
Kangle Li and Hoi~Chun Po.
\newblock ``Higher-dimensional jordan-wigner transformation and auxiliary
  majorana fermions''.
\newblock \href{https://dx.doi.org/10.1103/PhysRevB.106.115109}{Phys. Rev. B
  {\bf 106}, 115109}~(2022).

\bibitem{bochniak2020bosonization}
Arkadiusz Bochniak and B{\l}a{\.z}ej Ruba.
\newblock ``Bosonization based on {C}lifford algebras and its gauge theoretic
  interpretation''.
\newblock \href{https://dx.doi.org/10.1103/PhysRevD.102.114502}{Journal of High
  Energy Physics {\bf 2020}, 1--36}~(2020).

\bibitem{tantivasadakarn2020jordan}
Nathanan Tantivasadakarn.
\newblock ``{J}ordan-{W}igner dualities for translation-invariant
  {H}amiltonians in any dimension: Emergent fermions in fracton topological
  order''.
\newblock \href{https://dx.doi.org/10.1103/PhysRevResearch.2.023353}{Physical
  Review Research {\bf 2}, 023353}~(2020).

\bibitem{Carleo17}
Giuseppe Carleo and Matthias Troyer.
\newblock ``Solving the quantum many-body problem with artificial neural
  networks''.
\newblock \href{https://dx.doi.org/10.1126/science.aag2302}{Science {\bf 355},
  602--606}~(2017).

\bibitem{Choo18}
Kenny Choo, Giuseppe Carleo, Nicolas Regnault, and Titus Neupert.
\newblock ``Symmetries and many-body excitations with neural-network quantum
  states''.
\newblock \href{https://dx.doi.org/10.1103/PhysRevLett.121.167204}{Physical
  Review Letters {\bf 121}, 167204}~(2018).

\bibitem{wen2003quantum}
Xiao-Gang Wen.
\newblock ``Quantum orders in an exact soluble model''.
\newblock \href{https://dx.doi.org/10.1103/PhysRevLett.90.016803}{Physical
  review letters {\bf 90}, 016803}~(2003).

\bibitem{vicentini2021netket}
Filippo Vicentini, Damian Hofmann, Attila Szabó, Dian Wu, Christopher Roth,
  Clemens Giuliani, Gabriel Pescia, Jannes Nys, Vladimir Vargas-Calderón,
  Nikita Astrakhantsev, and Giuseppe Carleo.
\newblock ``{NetKet 3: Machine Learning Toolbox for Many-Body Quantum
  Systems}''.
\newblock \href{https://dx.doi.org/10.21468/SciPostPhysCodeb.7}{SciPost Phys.
  CodebasesPage~7}~(2022).

\bibitem{lu2019efficient}
Sirui Lu, Xun Gao, and L-M Duan.
\newblock ``Efficient representation of topologically ordered states with
  restricted boltzmann machines''.
\newblock \href{https://dx.doi.org/10.1103/PhysRevB.99.155136}{Physical Review
  B {\bf 99}, 155136}~(2019).

\bibitem{inui2021determinant}
Koji Inui, Yasuyuki Kato, and Yukitoshi Motome.
\newblock ``Determinant-free fermionic wave function using feed-forward neural
  networks''.
\newblock \href{https://dx.doi.org/10.1103/PhysRevResearch.3.043126}{Phys. Rev.
  Research {\bf 3}, 043126}~(2021).

\bibitem{jax2018github}
James Bradbury, Roy Frostig, Peter Hawkins, Matthew~James Johnson, Chris Leary,
  Dougal Maclaurin, George Necula, Adam Paszke, Jake Vander{P}las, Skye
  Wanderman-{M}ilne, and Qiao Zhang.
\newblock ``{JAX}: composable transformations of {P}ython+{N}um{P}y
  programs''~(2018).

\bibitem{hafner2021mpi4jax}
Dion H{\"a}fner and Filippo Vicentini.
\newblock ``mpi4jax: Zero-copy mpi communication of jax arrays''.
\newblock \href{https://dx.doi.org/https://doi.org/10.21105/joss.03419}{Journal
  of Open Source Software {\bf 6}, 3419}~(2021).

\bibitem{stokes2020phases}
James Stokes, Javier~Robledo Moreno, Eftychios~A Pnevmatikakis, and Giuseppe
  Carleo.
\newblock ``Phases of two-dimensional spinless lattice fermions with
  first-quantized deep neural-network quantum states''.
\newblock \href{https://dx.doi.org/10.1103/PhysRevB.102.205122}{Physical Review
  B {\bf 102}, 205122}~(2020).

\end{thebibliography}

\onecolumn
\appendix

\section{Code availability}
The exact reference energies and variational energies based on Jordan-Wigner transformation in Figs.~\eqref{fig:stretch1} and \eqref{fig:symmetry_results} were obtained using NetKet 3~\cite{vicentini2021netket}. The code to reproduce the experiments based on bosonization procedures are available upon request.

\section{Details of the bosonization procedure}\label{sec:bosonization}
We summarize the bosonization procedure from Refs.~\cite{po2021symmetric} for the case of the $t$-$V$-model model on a square lattice, and restrict the derivation to the even fermion sector. We introduce two Majorana fermion operators $\gamma^1$ and $\gamma^2$ with commutation relations $\{\gamma^i_{\vec{r}}, \gamma^j_{\vec{r}'}\} = 2\delta^{ij} \delta_{\vec{r},\vec{r}'}$:
\begin{align}
    f_{\vec{r}} = \frac{1}{2} (\gamma^1_{\vec{r}} - i \gamma^2_{\vec{r}}) ,  \qquad f^\dagger_{\vec{r}} = \frac{1}{2} (\gamma^1_{\vec{r}} + i \gamma^2_{\vec{r}})
\end{align}
In terms of the Majorana operators, the terms relevant for the Hamiltonian in Eq.~\eqref{eq:hamiltonian_fermion} are
\begin{align}
\begin{split}
    n_r &\to \frac{1}{2} (1 + i \gamma^2_{\vec{r}} \gamma^1_{\vec{r}} ) \\
    f^\dagger_{\vec{r}} f_{\vec{r}'} + f^\dagger_{\vec{r}'} f_{\vec{r}} &\to \frac{1}{2} (i \gamma^2_{\vec{r}} \gamma^1_{\vec{r}'} - i \gamma^1_{\vec{r}} \gamma^2_{\vec{r}'})
\end{split}
\end{align}

\subsection{Constructing bosonic operators}
We now introduce a set of auxiliary fermions in order to maintain locality in the fermion-qubit mapping.
We focus on a square lattice and introduce Majorana fermions $\eta^i$ ($i=1,2$) and $\chi^j$ ($j=1,2,3,4$), in order to separate the internal and spatial degrees of freedom, similar to the procedure in Ref.~\cite{kitaev2006anyons}. The $\eta^i$ modes capture the internal degrees of freedom, while the $\chi^j$ represent the spatial degrees of freedom. Each $\chi^j$ represents an edge attached to the lattice vertex $\vec{r}$. As shown in Fig.~\ref{fig:stretch1}, we associate a mode $\chi^j_{\vec{r}}$ to each of the edge attachments at site $\vec{r}$, and hence, the number of $\chi$ modes equals the coordination number. Since we consider square lattices, we require $4$ such modes. Furthermore, we only require two $\eta^i$ modes to describe spinless fermions.
We introduce a set of operators
\begin{align}
\begin{split}
    \Theta^{ij}_{\vec{r}} &= i \eta^i_{\vec{r}} \eta^i_{\vec{r}} \\
    \Lambda^{ij}_{\vec{r}} &= i \eta^i_{\vec{r}} \chi^i_{\vec{r}} \\
    \Upsilon^{ij}_{\vec{r}} &= -i \chi^i_{\vec{r}} \chi^j_{\vec{r}} \label{eq:bosonic_operators_def}
\end{split}
\end{align}
and focus on writing the Hamiltonian in Eq.~\eqref{eq:spinless_ham_boson} in terms of such operators.
We have the commutation relations
\begin{align}
    &\{\eta^i_{\vec{r}}, \eta^{j }_{\vec{r}'}\}=\{\chi^i_{\vec{r}}, \chi^{j }_{\vec{r}'}\} = 2\delta^{ij} \delta_{\vec{r},\vec{r}'} \label{eq:etachi_commutation}
\end{align}
Due to the commutation relations in Eq.~\eqref{eq:etachi_commutation}, we find that the operators $\Theta$, $\Lambda$, and $\Upsilon$ are bosonic, in the sense that they commute at different sites.

Since all terms are of the form $i \gamma^i_{\vec{r}} \gamma^j_{\vec{r}'}$, we will identify the (bosonic) operators in terms of the $\gamma_{\vec{r}}^i$ Majorana modes 
\begin{align}
    i \gamma^i_{\vec{r}} \gamma^j_{\vec{r}} &= \Theta^{ij}_{\vec{r}}  \label{eq:Theta}\\
    i \gamma^i_{\vec{r}} \gamma^j_{\vec{r}'} &= \Lambda^{i\mu}_{\vec{r}} \Lambda^{j\nu}_{\vec{r}'}  \label{eq:LambdaLambda}
\end{align}
Here, $\mu,\nu$ are numbers labeling the Majorana modes $\chi$ at two ends of the edge. In Eq.~\eqref{eq:LambdaLambda} we have implicitly inserted auxiliary operators $\chi^\mu_{\vec{r}} \chi^\nu_{\vec{r}'}$ terms in order to obtain pairs of operators at each lattice site, i.e.\ $\eta_{\vec{r}} \chi_{\vec{r}}$. This method is similar to, for example, the Verstraete-Cirac mapping~\cite{verstraete2005mapping} which explicitly inserts auxiliary operators between the original fermionic operators: $f_i^\dagger f_j \to f_i^\dagger (i c_{i'} d_{j'})f_j$, where $c = (b + b^\dagger)$ and $d = -i(b - b^\dagger)$ and $b^\dagger$ are auxiliary fermions. The result of this procedure is that Eqs.~\eqref{eq:Theta} and~\eqref{eq:LambdaLambda} only contain operator pairs at the same lattice sites and will therefore not produce non-local Jordan-Wigner strings when mapped onto qubit operators. 

\begin{figure}[tbh]
\centering
\includegraphics[width=0.35\textwidth, trim=41mm 13mm 40mm 15mm, clip]{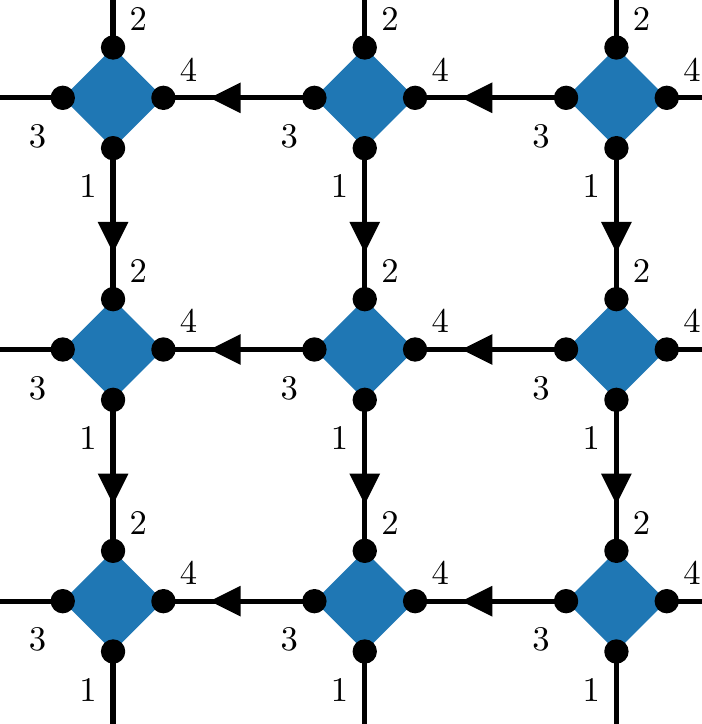}
\caption{Conventions for the ordering of the auxiliary $\chi^i_{\vec{r}}$ modes, and the directions of the edges that determine the sign of the $\Lambda^{i\mu}_{\vec{r}} \Lambda^{j\nu}_{\vec{r}'}$ terms. Blue boxes correspond to a single location vector $\vec{r}$.\label{fig:conventions}}
\end{figure}

The conventions with respect to the directions of the edges of our grid are depicted in Fig.~\ref{fig:conventions}, and determine the sign of Eq.~\eqref{eq:LambdaLambda}. 

Within this convention, we have the following identifications for the fermion operators
\begin{align}
\begin{split}
    n_{\vec{r}} &\to \frac{1}{2}(1+\Theta^{21}_{\vec{r}}) \\
    f^\dagger_{\vec{r}} f_{\vec{r}'} + f^\dagger_{\vec{r}'} f_{\vec{r}} &\to \frac{1}{2} (\Lambda^{2\mu}_{\vec{r}}\Lambda^{1\nu}_{\vec{r}'} - \Lambda^{1\mu}_{\vec{r}}\Lambda^{2\nu}_{\vec{r}'})\label{eq:subs_Lambda_munu}
\end{split}
\end{align}
and hence the Hamiltonian reads
\begin{align}
\begin{split}
    H =& -\frac{t}{2} \sum_{\vec{r}\in \mathcal{V}}  (\Lambda^{24}_{\vec{r}}\Lambda^{13}_{\vec{r}+\vec{x}} - \Lambda^{14}_{\vec{r}}\Lambda^{23}_{\vec{r}+\vec{x}} +
    \Lambda^{22}_{\vec{r}}\Lambda^{11}_{\vec{r}+\vec{y}} - \Lambda^{12}_{\vec{r}}\Lambda^{21}_{\vec{r}+\vec{y}}
    )  \\
    &- \frac{\mu}{2} \sum_{\vec{r} \in \mathcal{V}} (1 + \Theta^{21}_{\vec{r}}) \\
    &+ \frac{V}{4} \sum_{\vec{r} \in \mathcal{V}} \sum_{\delta\vec{r} \in \{\vec{x}, \vec{y}\}} (1 + \Theta^{21}_{\vec{r}})(1 + \Theta^{21}_{\vec{r}+\delta\vec{r}})
    \label{eq:spinless_ham_boson_operators}
\end{split}
\end{align}
which can be written in terms of $\eta$ and $\chi$ using definitions in Eq.~\eqref{eq:bosonic_operators_def}.

\subsection{Parton decomposition}
In a last step, we again move to fermionic (parton) operators. For the physical modes, we have
\begin{alignat}{2}
    \eta^1 &= a + a^\dagger , \qquad \eta^2 &= i(a - a^\dagger)
\end{alignat}
and for the auxiliary modes
\begin{alignat}{2}
    \chi^1 &= b + b^\dagger , \qquad \chi^2 &= i(b - b^\dagger) \nonumber\\
    \chi^3 &= c + c^\dagger , \qquad
    \chi^4 &= i(c - c^\dagger)
\end{alignat}

Notice that our parton Hilbert space is now $2^3$ dimensional at each site.
The on-site Hilbert space reads
\begin{align}
\ket{n_a n_b n_c} = (a^\dagger)^{n_a} (b^\dagger)^{n_b}  (c^\dagger)^{n_c} \ket{\Omega}
\end{align}
where $n_{i=a,b,c} \in \{0,1\}$ and $\ket{\Omega}$ represents the vacuum.

\subsection{On-site Jordan-Wigner transformation}
The bosonic operators defined in Eq.~\eqref{eq:bosonic_operators_def} commute on different sites, i.e.\ for $\vec{r} \neq \vec{r}'$, we have $[\Delta_{\vec{r}}, \Delta'_{\vec{r}'}] = 0$ for all combinations of $\Delta, \Delta' \in \{\Theta, \Lambda, \Upsilon\}$. 
Since we have written the Hamiltonian in terms of bosonic operators in Eq.~\eqref{eq:spinless_ham_boson_operators} we do need to carry out a Jordan-Wigner transformation for fermions at different lattice sites. However, since the operators in Eq.~\eqref{eq:bosonic_operators_def} do not commute on the same site, we can most easily identify the spin representations of the bosonic operators by carrying out a Jordan-Wigner transformation of their parton components on each site in order to reproduce the correct anti-commutation relations. Since we have $3$ fermion modes $\{a^\dagger, b^\dagger, c^\dagger\}$, we require $3$ qubits per site
\begin{align}
\begin{split}
    a^\dagger_{\vec{r}} &\to Q^{(1)-}_{\vec{r}}  \\
    b^\dagger_{\vec{r}} &\to Z^{(1)}_{\vec{r}}Q^{(2)-}_{\vec{r}}  \\
    c^\dagger_{\vec{r}} &\to Z^{(1)}_{\vec{r}}Z^{(2)}_{\vec{r}} Q^{(3)-}_{\vec{r}} \label{eq:jw}
\end{split}
\end{align}
where we identified $\ket{0} = \ket{\uparrow}$ and $\ket{1} = \ket{\downarrow}$ and
\begin{align}
Q^{(i)-}_{\vec{r}} = \frac{X^{(i)}_{\vec{r}} - i Y^{(i)}_{\vec{r}}}{2}
\end{align}
As we will show in the next section, using the abovementioned mapping, one can rewrite the Hamiltonian in Eq.~\eqref{eq:spinless_ham_boson_operators} in terms of Pauli operators using the definitions in Eq.~\eqref{eq:bosonic_operators_def}. However, we first demonstrate how to eliminate the effect of one of the auxiliary qubits in Eq.~\eqref{eq:jw}.

\subsection{Constraints}
The Hilbert space is enlarged due to the $b^\dagger$ and $c^\dagger$ auxiliary fermion modes. The on-site parity operator $\Gamma_{\vec{r}}$ for any site commutes with the Hamiltonian in Eq.~\eqref{eq:spinless_ham_boson}. To reduce the Hilbert space, we restrict the on-site parton parity operator (which commutes with the Hamiltonian). 
\begin{align}
\Gamma_{\vec{r}} &= (-i)^{\abs{\mathcal{M}_{\vec{r}}}/2} \prod_{g \in \mathcal{M}_{\vec{r}}} g \overset{c}{=} 1
\end{align}
% \textbf{In the above, might be that $i \to -i$}
where $\mathcal{M} = \{\eta^i\}_{i=1,2}\cup \{\chi^j\}_{j=1,2,3,4}$ is the set of Majorana fermions.
In the spin basis, we find
\begin{align}
    \Gamma = -Z^{(1)}Z^{(2)} Z^{(3)} \overset{c}{=} 1
\end{align}
Hence, the third spin is redundant and we can remove it by absorbing its effect in the other spins
\begin{align}
\begin{split}
    X^{(3)} &\to 1 \\
    Z^{(3)} &\to - Z^{(1)}Z^{(2)} \\
    Y^{(3)} &\to -i Z^{(1)}Z^{(2)} \label{eq:constraint1}
\end{split}
\end{align}
Using Eq.~\eqref{eq:constraint1} to eliminate the third qubit, we obtain the expression for the Hamiltonian in Eq.~\eqref{eq:spinless_ham_boson}.

\subsection{Constraints}
We still have redundant degrees of freedom. To see this, we write the identity operator in terms of fermionic operators on the boundary of a plaquette attached to $\vec{r}$
\begin{align}
\begin{split}
    &(i \gamma^1_{\vec{r}} \gamma^2_{\vec{r}}) (i \gamma^2_{\vec{r}} \gamma^1_{\vec{r}+\vec{x}}) (i \gamma^1_{\vec{r}+\vec{x}} \gamma^2_{\vec{r}+\vec{x}})
    (i \gamma^2_{\vec{r}+\vec{x}} \gamma^1_{\vec{r}+\vec{x}+\vec{y}}) \\
    &\times (i \gamma^1_{\vec{r}+\vec{x}+\vec{y}} \gamma^2_{\vec{r}+\vec{x}+\vec{y}}) (i \gamma^2_{\vec{r}+\vec{x}+\vec{y}} \gamma^1_{\vec{r}+\vec{y}}) \\
    &\times (i \gamma^1_{\vec{r}+\vec{y}} \gamma^2_{\vec{r}+\vec{y}}) (i \gamma^2_{\vec{r}+\vec{y}} \gamma^1_{\vec{r}}) \\
    &= \iden
\end{split}
\end{align}
While the above reduces to the identity operator in the fermionic formalism, it introduces a (gauge) constraint on the bosonic side. We can rewrite the above in terms of bosonic operators using Eq.~\eqref{eq:Theta}--\eqref{eq:LambdaLambda}, and ultimately in terms of the gauge operators $\Upsilon$ using Eq.~\eqref{eq:bosonic_operators_def}. After some algebra, we obtain the constraint 
\begin{align}
    \Upsilon^{24}_{\vec{r}} \Upsilon^{32}_{\vec{r}+\vec{x}} \Upsilon^{13}_{\vec{r}+\vec{x}+\vec{y}} \Upsilon^{41}_{\vec{r}+\vec{y}} &\overset{c}{=} -1 \label{eq:constraint_upsilon}
\end{align}
Hence, the auxiliary system is subject to a Gauss-law constraint of the form 
\begin{align}
    G_{\vec{r}} &\overset{c}{=} 1 \qquad \forall \vec{r} \in \mathcal{V}%\label{eq:gausslaw}
\end{align}

Imposing the boundary conditions in the fermionic system, we obtain the additional constraints introduced by non-contractable Wilson loops $W_{x, y}$: $W_{x, y} = (-1)^{L_{x, y}}$. To see this, we carry out the same procedure as in Eq.~\eqref{eq:constraint_upsilon} on these loops (for even-by-even tori). After bosonisation, we obtain the following spin operator identities that need to be satisfied
\begin{align}
     \oprod_{m=0}^{L_x-1} \Upsilon^{34}_{\vec{r}+m\vec{x}} \overset{c}{=} -1 &\to \oprod_{m=0}^{L_x-1} Z^{(1)}_{\vec{r}+m\vec{x}} Z^{(2)}_{\vec{r}+m\vec{x}} \overset{c}{=} -1 \\%\label{eq:pbc_constraint_x} \\      
     \oprod_{m=0}^{L_y-1} \Upsilon^{12}_{\vec{r}+m\vec{y}} \overset{c}{=} -1 &\to \oprod_{m=0}^{L_y-1} Z^{(2)}_{\vec{r}+m\vec{y}} \overset{c}{=} -1 %\label{eq:pbc_constraint_y}
\end{align}

\section{Symmetries}\label{sec:symmetry}
Rotations of $\pi/2$ are decomposed as
\begin{align}
    C_4 = V_{C_4} C_4^b
\end{align}
where $C_4^b$ rotates the grid, and $V_{C_4}$ handles the rotation of the $\chi$ degrees of freedom.
\begin{align}
    V_{C_4, \vec{r}} &= \frac{1- Z^{(1)}_\vec{r}}{2} \frac{1+iZ^{(2)}_{\vec{r}}}{\sqrt{2}} \\
    &+ \frac{1+ Z^{(1)}_\vec{r}}{2} iX^{(2)}_{\vec{r}} \frac{1-iZ^{(2)}_{\vec{r}}}{\sqrt{2}}
\end{align}

We again try to find the operation $U_{M} = V_M M^b$, where $M$ is a mirror operator. For $M_x$ (which flips the x-axis, i.e.\ $M_x(x, y) \to M_x(-x, y)$)
\begin{align}
    M_x &= V_{M_x} M_x^b 
\end{align}
where
\begin{align}
    V_{M_x} = \oprod_{\vec{r} \in \mathcal{V}} \frac{1-i Z_\vec{r}^{(1)}Z_\vec{r}^{(2)}}{\sqrt{2}}
\end{align}
and similarly
\begin{align}
    V_{M_y} = \oprod_{\vec{r} \in \mathcal{V}} \frac{1+iZ_\vec{r}^{(2)}}{\sqrt{2}}
\end{align}

For translation, we find $V_{T_{x,y}} = 1$, such that $U_{T_{x,y}} = T^b_{x,y}$.

% \appendix
\section{Optimization details}
We use variational Markov-Chain Monte Carlo (MCMC) with stochastic gradient descent (SGD) to optimize the wave function. 
We run the optimization for $12.5k$ steps, with $4096$ Monte Carlo  samples at each step. We run between $16$-$64$ MCMC walkers in parallel. The gradients are determined through automatic differentiation of the RBM wave function and decorrelated with Stochastic Reconfiguration~\cite{Carleo17}. The model implementation uses NetKet 3~\cite{vicentini2021netket}, based on Jax~\cite{jax2018github}, and uses MPI4Jax~\cite{hafner2021mpi4jax} to run MCMC walkers in parallel.

\section{Open boundary conditions}\label{sec:obc}

To obtain Eq.~\eqref{eq:spinless_ham_boson}, we assumed periodic boundary conditions. Similarly, one can consider open-boundary conditions, where special care must be taken for sites at the boundary. Compared to the bulk, the edge attachement number of boundary sites are lower. Therefore, one needs fewer spins to represent the corresponding site modes. Hence, one can follow a similar derivation as for periodic boundary conditions and decouple unused spins~\cite{po2021symmetric}. Notice that we have made our methodology generalizable for other forms of the constraints. For example, for the Gauss constraint, our methods only requires that it is of the form in Eq.~\eqref{eq:gausslaw}.

\section{Odd fermion sector}\label{sec:odd_fermions}

In our derivation and experiments, we have considered an even number of fermions $N_f$. In this sector, the bosonized operators remain local, and we are not hindered by the effect of Jordan-Wigner strings along the chosen Jordan-Wigner ordering.
To describe the odd-parity sector of the Hilbert space, however, one must introduce a defect in the lattice in Fig.~\ref{fig:conventions} and reverse the direction of one edge (see detailed discussion in Ref.~\cite{po2021symmetric}). In this regime, the defect introduces similar effects as a Jordan-Wigner string which reaches from the defect to the position of an odd-fermion operator. 

\section{Scaling behavior}\label{sec:scaling}

Using the local fermion-to-qubit mappings, one obtains a local Hamiltonian. Therefore, every Hamiltonian operator term can be evaluated in constant time, compared to JWT-based methods that scale with the size of the Jordan-Wigner strings, i.e.\ $\order{\sqrt{N}}$. These strings appear for $\order{\sqrt{N}}$ terms in the Hamiltonian.

Another part of the computation is evaluating the model for the probability amplitudes.
Sections~\ref{sec:solution_reduce} and~\ref{sec:doubly_canonical} require a computational time of $\order{N}$ per sample $\vec{\sigma}$. Preparing the vacuum state in Eq.~\eqref{eq:vacuum} can similarly be carried out in $\order{N}$ time. While in principle the vacuum state needs to be constructed only once in the method of Section~\ref{sec:vacuum_reduction}, scalability requires it to be computed on the fly for each sample. Furthermore, the bosonized operator in Eq.~\eqref{eq:xi3} introduces strings between pairs of fermion creation operators that can be computed on any path between them, including the shortest path (since the Gauss constraint is satisfied, the chosen path is irrelevant, in contrast to Jordan-Wigner transformations). From these considerations, we also find that the method in Section~\ref{sec:vacuum_reduction} scales as $\order{N}$ per sample. This can be compared to JWT-based methods, where the fermionc anti-symmetry requirements are fulfilled by the Jordan-Wigner strings in the Hamiltonian (see the discussion above).  Additionally, our method can be compared to the first-quantization approach studied e.g.\ in Ref.~\cite{stokes2020phases}, where determinants enforce the fermionic anti-symmetry properties. If we assume $\order{N_f}=\order{N}$ (e.g.\ at half filling), such a method scales as $\order{N^3}$ per sample.

\section{Pseudocodes}

\begin{algorithm}[H]
\caption{Generate initial random states\label{alg:randomstates}}
\begin{algorithmic}
\Require A fixed number of fermions $N_f$.
\State $\vec{\sigma}^{(1)} \gets $ List of size $N$ with $N_f$ $1$'s and $N-N_f$ $(-1)$'s
\State $\vec{\sigma}^{(1)} \gets $ Permute($\vec{\sigma}^{(1)}$)\hspace*{\fill} (see Eq.~\eqref{eq:Nf_constraint})
\State $\vec{\sigma}^{\textrm{sub}} \gets (L_x-1)\times(L_y-1)$ random spins $\in \{-1, 1\}$
\State $\vec{\sigma}^{x} \gets -\prod_{m=0}^{L_y-2} \sigma^{\textrm{sub}}_{\vec{r}+m\vec{y}}$\hspace*{\fill} (see Eq.~\eqref{eq:pbc_constraint_x})
\State $\vec{\sigma}^{y} \gets -\prod_{m=0}^{L_x-1} \sigma^{(1)}_{\vec{r}+m\vec{x}} \prod_{m=0}^{L_x-2} \sigma^{\textrm{sub}}_{\vec{r}+m\vec{x}}$ \hspace*{\fill} (see Eq.~\eqref{eq:pbc_constraint_y})
\State $\vec{\sigma}^{(2)} \gets $ Concatenate($\vec{\sigma}^{\textrm{sub}}$, $\vec{\sigma}^x$, $\vec{\sigma}^y$) 
\end{algorithmic}
\end{algorithm}

\begin{algorithm}[H]
\caption{Canonical sample reduction}\label{alg:canred}
\begin{algorithmic}
\Require A sample $\vec{\sigma} = (\vec{\sigma}^{(1)}, \vec{\sigma}^{(2)})$ and a (zizag) ordering of the sites $O = (\vec{r}_1, ..., \vec{r}_N)$.
\State $\xi \gets 1$
\State $\vec{\alpha} \gets \vec{\sigma}$
\For{$\vec{r}$ in $O$}
    \State $m_{\vec{r}} \gets \alpha_{\vec{r}}$
    \If{$m_{\vec{r}} = 1$}
        \State $\mu' \ket{\vec{\alpha}'} \gets G_{\vec{r}}\ket{\vec{\alpha}}$
        \State $\xi \gets \mu'.\xi $
        \State $\ket{\vec{\alpha}} \gets \ket{\vec{\alpha}'}$
    \EndIf
\EndFor
\end{algorithmic}
\end{algorithm}

\end{document}